\documentclass[12pt]{article}

\usepackage[english]{babel}

\usepackage{amsmath}
\usepackage{epsf}
\usepackage{epsfig}
\usepackage{amssymb}
\usepackage{graphicx}
\usepackage{multirow} 

\def\u{\tilde{u}}
\def\s{\tilde{s}}

\def\P{\mathcal{P}}
\def\u{\tilde{u}}
\def\s{\tilde{s}}

\newcommand\ainv{A'\to invisible}

\newcommand\g{\gamma}
\newcommand\ma{m_{A'}}

\newcommand\na{{n}_{A'}}

\begin{document}

\title{The exact tree-level calculation of the  dark photon production in high-energy electron scattering at the CERN SPS}

\author{  S.N.~Gninenko$^{1}$\footnote{{\bf e-mail}: sergei.gninenko@cern.ch},
 D.V.~Kirpichnikov$^{1}$\footnote{{\bf e-mail}: kirpich@ms2.inr.ac.ru},
   M.M.~Kirsanov$^{1}$\footnote{{\bf e-mail}: mikhail.kirsanov@cern.ch}
    and N.V.~Krasnikov$^{1,2}$\footnote{{\bf e-mail}: nikolai.krasnikov@cern.ch}
\\
\\
 $^{1}$ Institute for Nuclear Research of the Russian Academy of Sciences, \\117312 Moscow, Russia \\
 $^{2}$ Joint Institute for Nuclear Research, 141980 Dubna, Russia}


\date{\today}

\maketitle

\begin{abstract}

 Dark photon ($A'$) that couples to the standard model fermions via the kinetic mixing with photons  and serves as a mediator
of dark matter production could be observed in the high-energy electron scattering  $e^- + Z ~\rightarrow e^- + Z + A'$ off nuclei followed by 
the  $A' \to invisible $ decay. We have performed the exact  tree-level 
calculations of the $A'$ production cross sections and implemented them in the program for the full simulation
of such events in the experiment NA64 at the CERN SPS. Using simulations results, 
we study the missing  energy signature for the bremsstrahlung $\ainv$  decay that permits the determination of the $\g-A'$ mixing strength
in a wide, from sub-MeV to sub-GeV, $A'$ mass range. We refine and expand our earlier studies of this signature for
discovering $A'$ by including corrections to the previously used calculations based on the improved  Weizsaker-Williams approximation,  
which turn out to be significant. We compare our cross sections values with the results from other calculations and find a good agreement
between them. The possibility of future measurements with high-energy electron beams and the sensitivity to $A'$ are briefly discussed.
\end{abstract}

\section{Introduction}

At present there are several  indications  that  new physics beyond the Standard Model (SM) exists. The most robust fact is the existence of
dark matter (DM), see e.g. \cite{Universe1} -\cite{Universe3}.  Among  many models of DM, there is a class of models that motivate   
the existence of  light DM with a mass $ \lesssim 1$ GeV, for a review see e.g.  \cite{lightdark,lightdark1}. 
In such  renormalizable models,  there must be  a new light vector(scalar) boson   which serves as a   mediator 
connecting   the visible and  DM particles \cite{lightdark1}. 
The most popular realisation of the light DM paradigm is the model with a massive vector boson, called
dark photon ($A'$) which interacts with the Standard Model (SM) particles through the mixing with the ordinary photon.
The possibility to observe $A'$ through its visible (into SM particles) or invisible (into light DM particles) decay
in accelerator experiments motivate significant efforts devoted to searches for dark photons, for a review see e.g. \cite{lightdark1}. 

The fixed-target experiment NA64 at the CERN SPS \cite{Gninenko:2013rka,Andreas:2013lya,Gninenko:2016kpg,na64prl,na64prd} is designed to search for
$A'$ that can be produced through the mixing with the bremsstrahlung photon from the reaction of high-energy electron scattering off nuclei
\begin{equation}
e^- + Z ~\rightarrow e^- + Z + A' \,
\label{reaction}
\end{equation} 
 accompanied by the dominant invisible $A'$ decay. The discussions of the missing energy signature of this process and the distinctive  
distributions of variables that can be used to distinguish the $\ainv$ signal from background including 
the results of the detailed simulation of the detector response to events with and without $A'$ emission and estimates of 
the efficiency of the signal event selection can be found in Ref.~\cite{Gninenko:2016kpg}.
The evaluation of the sensitivity of the experiment shows that it can
probe the still unexplored area of the mixing strength $10^{-6}\lesssim \epsilon \lesssim 10^{-2}$ and masses up to $M_{A'} \lesssim 1$ GeV.
In the case of  the signal observation, a possibility of determining the $A'$ mass $m_{A'}$ and mixing strength
$\epsilon$ using the missing energy spectrum shape was also discussed.

The results presented in Ref. \cite{Gninenko:2016kpg, na64prl} were based on the full Monte Carlo simulations  obtained by using the improved 
Weizsaker-Williams (IWW) approximation  \cite{Tsai:1973py, Bjorken:2009mm}
for the differential cross section calculations of the process \eqref{reaction}. Recently, the exact, tree-level (ETL) calculations of the cross
sections for the $A'$ production 
in the reaction \eqref{reaction} performed by Liu et al. were published \cite{Liu:2017htz}. It was pointed out that for a certain
range of the initial electron beam energy $E_0$ and the $A'$ mass $\ma$, the $A'$ yield calculated in the IWW framework could differ
significantly from the one calculated exactly at tree-level \cite{Liu:2017htz}.
It is clear that reliable theoretical predictions for the $A'$ yield are essential 
for a proper interpretation of the experimental results both in case of observation of signal or non-observation.
In the latter case they are necessary to obtail robust exclusion limits in the $A'$ parameter space. 
Therefore, it is instructive to perform an accurate calculation of the $A'$ cross-section  based on precise phase space integration over
the final state  particles in the reaction  $e^- Z\rightarrow e^-ZA'$. This provides the main motivation for this work.
First, it is crucial to cross-check and confirm the results of Ref. \cite{Liu:2017htz}, which could be also useful for
other dark force experiments planned with electron beams \cite{lightdark1}. Second, it is important to implement 
the  ETL cross sections in the NA64 simulatin software.

The organisation of the paper is as follows. In Section 2 we present the results 
for ETL calculations  of the cross section relevant for the NA64 experiment 
and compare them with analogous calculations based on the use of the IWW approximation and 
with the results of Liu et al. \cite{Liu:2017htz}.
In Section 3 we implement  ETL cross sections for the reaction $e^- + Z ~\rightarrow e^- + Z + A'$ 
in the NA64 simulation software and compare obtained results with the previous full simulation calculations
based on the use of the IWW approximation. In Section 4 we evaluate the sensitivity in terms of the mixing parameter $\epsilon$
of the NA64 experiment to invisibly decaying dark photons in the wide range of $A'$ masses. Section 5 contains concluding remarks.

\section{Cross sections}

The Lagrangian of 
the dark sector has the following form 
\begin{eqnarray}
\mathcal{L} = -\frac14 F'_{\mu\nu}F'^{\mu\nu}+
\frac{\epsilon}{2}F'_{\mu\nu}F^{\mu\nu}+  \frac{m_{A'}^2}{2}A'_\mu A'^\mu \nonumber \\
+ i \bar{\chi}\gamma^\mu \partial_\mu \chi- 
m_\chi \bar{\chi} \chi -
e_D \bar{\chi}\gamma^\mu A'_\mu \chi, 
\label{DarkSectorLagrangian}
\end{eqnarray}
where  $A'_\mu$ is a massive Dark photon field which originates from 
 broken $U'(1)$  gauge group,
 $\epsilon$ is the value of the  kinetic mixing between SM photon and Dark photon, $\chi$ is Dirac spinor which can be treated as Dark 
Matter fermions  coupled to 
$A'_{\mu}$ by the 
  coupling constant  $e_D$. The interaction between the
  Dark photon and SM fermions are described by the following Lagrangian
\begin{equation}
\mathcal{L}_{int}= \epsilon e A'_{\mu} J^\mu_{em}.
\end{equation}
The decay rates of  invisible   and visible   decays   
 are given by
 \begin{eqnarray}
\Gamma (A'\rightarrow \bar{\chi}\chi) = \frac{\alpha_D}{3} m_{A'}\bigl(1+
\frac{2m_\chi^2}{m_{A'}^2}\bigr)\sqrt{1-\frac{4m_\chi^2}{m_{A'}^2}}, \qquad \nonumber \\  
\Gamma (A'\rightarrow  e^- e^+) = \frac{\alpha_{QED} \epsilon^2}{3} m_{A'}\bigl(1+
\frac{2m_e^2}{m_{A'}^2}\bigr)\sqrt{1-\frac{4m_e^2}{m_{A'}^2}},
\end{eqnarray}
where $\alpha_{QED} = e^2/4\pi$, $\alpha_D = e_D^2/4\pi$ and 
$m_e, ~m_\chi$ are the electron and DM particle masses, respectively.
We suppose that the $A'$ lepton decay channel is suppressed,
$\Gamma (A'\rightarrow \bar{\chi}\chi) \gg \Gamma (A'\rightarrow  e^- e^+)$, and 
that dark matter invisible decay mode is dominant, i.e. 
$\Gamma (A'\rightarrow \bar{\chi}\chi)/\Gamma_{tot} \simeq 1$.

\subsection{The IWW cross sections}
\label{MCsection}


In the   IWW approximation  the 
differential and total 
cross sections for the reaction (1)  for $m_{A'} \gg m_e$ 
can  be written as ~\cite{Bjorken:2009mm}
\begin{equation}
\frac{d\sigma_{\text{WW}}^{A'}}{dx} =  (4\alpha^3 \epsilon^2 \chi)(m^2_{A'}\frac{1-x}{x} +m^2_ex)^{-1} \times 
(1 - x + \frac{x^2}{3}) \,,
\label{diff-crsec}
\end{equation}
\begin{equation}
\sigma_{\text{WW}}^{A'} = \frac{4}{3} \frac{\epsilon^2 \alpha^3 \chi }{m_{A'}^2} \cdot \log \delta^{-1}_{A'}, \qquad 
\delta_{A'} = \mbox{max} \left[ \frac{m_e^2}{m_{A'}^2}, \frac{m_{A'}^2}{E_0^2}\right],
\label{TotIWWcrSect1}
\end{equation}
 where
$\chi$ is an effective flux of photons
\begin{equation}
\chi   = \int^{t_{max}}_{t_{min}} dt \frac{(t - t_{min})}{t^2}\left[G^{\text{el}}_2(t)+G^{\text{inel}}_2(t)\right] \,.
\end{equation}
and $x = \frac{E_{A'}}{E_o}$.
Here $t_{min} = m_{A'}^4/(4 E_0^2)$, $t_{max}=m_{A'}^2$ and $ G^{\text{el}}_2(t)  $, $  G^{\text{inel}}_2(t) $ are elastic and 
inelastic form-factors respectively. 
For usual energies the elastic form-factor $ G^{\text{el}}_2(t)  $  dominates.
The elastic form-factor can be written as
\begin{equation}
G^{\text{el}}_2(t)  = \left(\frac{a^2t}{1 + a^2 t}\right)^2 \left(\frac{1}{1 +t/d}\right)^2  Z^2 \,,
\end{equation}
where $ a = 111 Z^{-1/3}/m_e$ and
$d = 0.164~GeV^2~A^{-2/3}$. In our case we consider the quasi-elastic  reaction \eqref{reaction}, so 
that inelastic nuclear formfactor is not taken into account. 
Numerically, $\chi = Z^2 \cdot Log$, where the function
$Log$ weakly depends on atomic screening, nuclear size effects and kinematics \cite{Tsai:1973py,Bjorken:2009mm}.
For $m_{A'} \leq 500~MeV$ $Log \approx (5 - 10)$.

\subsection{The ETL cross sections}
In this subsection we closely follow  
notations of Ref.~\cite{Liu:2017htz} to derive the  production 
cross-section of the $A'$ beyond the IWW approximation. 
To begin with we consider kinematic variables of the  process 
$$
e^-(p)+Z(P_i)\rightarrow e^-(p')+N(P_f)+A'(k),
$$
here $p=(E_0, {\bf p})$ is the 4-momentum of incoming electron, 
$P_i= (M, 0)$ denotes the Z Nucleus 4-momentum in the initial state, 
final state Z Nucleus momentum is defined by $P_f= (P^0_f, {\bf P}_f )$,
the $A'$-boson 
momentum is $k=(k_0, {\bf k})$ and $p'=(E', {\bf p}')$ is the 
momentum of electron recoil. It is convenient  
to perform calculation in the frame where vector ${\bf V = p-k}$ is 
parallel to $z$-axis and vector ${\bf k}$ is  in the  $xz$-plane. 
We define a 4-momentum  transfer to the nucleus as $q=P_i -P_f$. In
this frame the polar and axial angles of ${\bf q}$  
are denoted by  $\theta_q$ and $\phi_q$ respectively. 
In contrast to Ref.~\cite{Liu:2017htz}
 we use mostly minus metric, such that 
$t=-q^2=|{\bf q}|^2-q_0^2>0$. 
 After some algebraic manipulations one can obtain
\begin{equation}
\cos \theta_q = 
-\frac{|{\bf V}|^2+|{\bf q}|^2+m_e^2-(E_0+q_0-k_0)^2}{2|{\bf V}| |{\bf q}|}, 
\qquad q_0 =-\frac{t}{2M}, 
\qquad |{\bf q}|= \sqrt{\frac{t^2}{4M^2}+t}.
\end{equation} 
We assume that nucleus has zero 
 spin~\cite{Liu:2017htz,Liu:2016mqv,Beranek:2013nqa,Beranek:2013yqa}, 
 as a consequence  the photon-nucleus vertex is given by  
$$i e \sqrt{G_2^{\text{el}}(t)} (P_i+P_f)_\mu \equiv i e \sqrt{G_2^{\text{el}}(t)} \P_\mu.$$
We also use  a
convenient designation for the partial energy of $A'$-boson $x=k_0/E_0$ 
and for the angle $\theta$ between ${\bf k}$ and ${\bf p}$.


\begin{figure}[h]
\begin{center}
 \includegraphics[width=0.45\textwidth,height=0.48\textwidth]{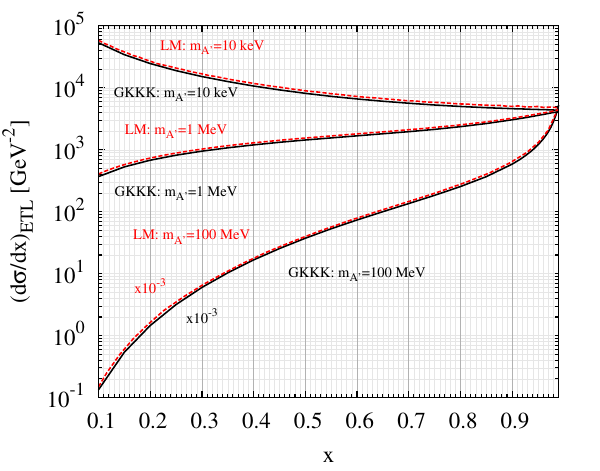}
\includegraphics[width=0.543\textwidth]{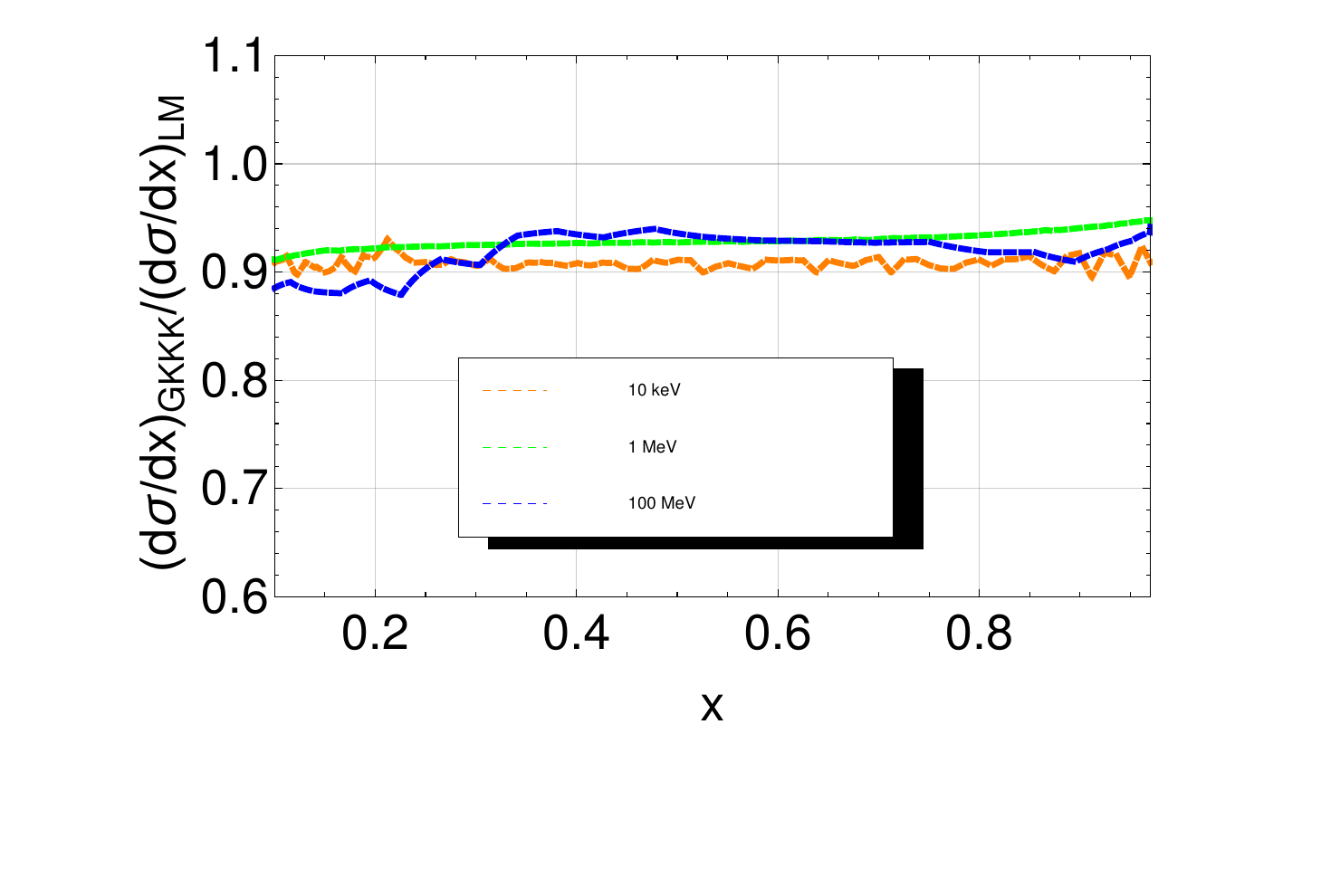}
\caption { Left panel: Comparison of the values of the differential cross 
section $(d\sigma/dx)_{ETL}  $  as a function of $x = E_{A'}/E_0$  for 
various $A'$ masses  and mixing strength $\epsilon = 1$ calculated in 
this work (curves labeled with GKKK) and  in Ref. \cite{Liu:2017htz} (LM 
curves).  
The spectra are normalized to the same number of  electron on target.
Right panel: ratio of ETL cross-sections shown on the left panel for 
various masses of dark photon, $10$ keV $<m_{A'}<100$ MeV.
\label{crsec}}
\end{center}
\end{figure}

\begin{figure}[h]
\begin{center}
\includegraphics[width=0.6\textwidth]{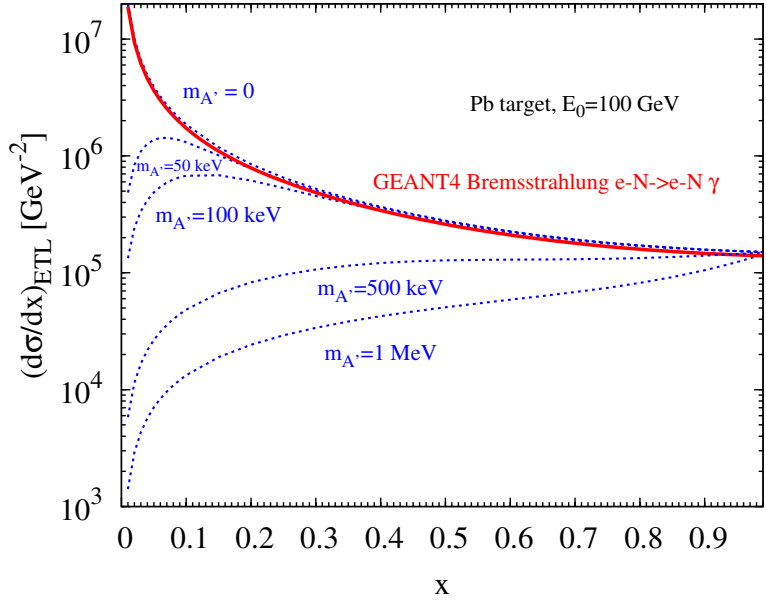}
\caption { Differential cross-section of $A'$ production   as a function of  $x$ for various masses $m_{A'} < 1$~MeV and $\epsilon = 1$.}
\label{figure7}
\end{center}
\end{figure}


One can express the relevant differential cross-section 
in the following form~\cite{Liu:2017htz}
\begin{equation}
\frac{d\sigma}{dx \, d\cos \theta} =
 \frac{\epsilon^2  \alpha^3 |{\bf k}|E_0 }{|{\bf p}|
|{\bf k - p}|  } \cdot  \int\limits_{t_{min}}^{t_{max}} \frac{d t}{t^2}
\, G_{2}^{el}(t) \cdot  \int\limits_{0}^{2\pi} \frac{d\phi_q}{2\pi}
\, \frac{|A^{2\rightarrow 3}_{X}|^2}{8 M^2}, 
\label{AprDiff1}
\end{equation}
where $t_{\text{min}}$ and $t_{\text{max}}$ are the values of minimum and 
maximum  momentum transfer respectively. The quantities for 
$t_{\text{min}}$ and $t_{\text{max}}$
 are derived explicitly in Ref.~\cite{Liu:2017htz}. The
 production amplitude squared for dark photon is 
 given by
$$
|A^{2\rightarrow 3}_{A'}|^2=
\frac{2}{\s^2 \u^2}
\Bigl(
2 m_e^2 (\P^2 t (\s+\u)^2-4((\P \cdot p) \u +(\P \cdot p') \s)^2)+
m_{A'}^2 (\P^2 t (\s-\u)^2-4((\P \cdot p) \u +(\P \cdot p') \s)^2)+
$$
\begin{equation} 
+\s \u (\P^2 ((\s +t)^2+(\u+t)^2) -4 t[( \P \cdot p)^2 +( \P \cdot p')^2 ])\Bigr),
\end{equation}
where the mandelstam variables and relevant dot products are 
\begin{equation}
\s =(p'+ k)^2-m_e^2 = 2(p'\cdot k)+m_{A'}^2, \qquad
\u =(p - k)^2-m_e^2 = - 2(p\cdot k)+m_{A'}^2,
\end{equation}
\begin{equation}
\P^2 = 4 M^2+t, \qquad \P \cdot p = 2M E_0 k_0 -(\s+t)/2, \qquad
\P \cdot p' = 2M(E_0-k_0)+(\u-t)/2.
\end{equation}
We performed the integration of $|A_{A'}^{2 \rightarrow 3 }|^2$ over $\phi_q$ and found  
that
\begin{equation}
\int\limits_{0}^{2\pi} \frac{d\phi}{2\pi}
\, |A^{2\rightarrow 3}_{A'}|^2= A^{(0)}_{A'}+Y\cdot A^{(1)}_{A'}+\frac{1}{W^{1/2}} \cdot A^{(-1)}_{A'}+\frac{Y}{W^{3/2}} \cdot A^{(-2)}_{A'},
\end{equation} 
The relevant functions $A^{(i)}_{A'}$ are given by 
\begin{equation}
A^{(-2)}_{A'}=-8 M \left(4 E_0^2 M-t \left(2 E_0+M\right)\right) \left(m_{A'}^2+2 m_e^2\right), \qquad A^{(1)}_{A'}= \frac{8 M^2}{\tilde{u}},
\end{equation} 
\begin{equation}
A^{(-1)}_{A'}=\frac{8}{\tilde{u}} \Bigl[ M^2 \left(2 t \tilde{u}+\tilde{u}^2+4 E_0^2 \left(2 (x-1) \left(m_{A'}^2+2 m_e^2\right)-t ((x-2) x+2)\right)+2 t \left(-m_{A'}^2+2 m_e^2+t\right)\right)-
\end{equation}
$$
-2 E_0 M t
   \left((1-x) \tilde{u}+(x-2) \left(m_{A'}^2+2 m_e^2+t\right)\right)+t^2 \left(\tilde{u}-m_{A'}^2\right)\Bigr],
$$
\begin{equation}
A^{(0)}_{A'}= \frac{8}{\tilde{u}^2} \Bigl[ M^2 \left( 2t \tilde{u} + (t-4 E_0^2 (x-1)^2) \left(m_{A'}^2+2
   m_e^2\right)\right)+2 E_0 M t \left(\tilde{u}-(x-1) \left(m_{A'}^2+2
   m_e^2\right)\right) \Bigr],
\end{equation}

The functions  $Y$ and $W$  have the form
\begin{equation}
Y = q^2+2 q_0 E_0 - 2 \frac{|{\bf q}| |{\bf p}|}{|{\bf V}|} 
(|{\bf p}|-|{\bf k}| \cos \theta) \cos \theta_q, 
\quad 
W = Y^2 - 4  \frac{|{\bf q}|^2 |{\bf p}|^2 |{\bf k}|^2}{|{\bf V}|^2} 
 \sin^2 \theta \sin^2 \theta_q.
\end{equation}  
To derive the total cross-section  
we integrate (\ref{AprDiff1}) over all kinematically allowed variables $x$ and $\theta$.

In the numerical estimates we use the electron energy $E~=~100$~GeV and 
the lead target with $Z = 82$ relevant for the NA64 experiment.
In Fig.\ref{crsec} we present the results of our numeric 
calculations for the differential and total cross sections. 
In the left panel of Fig. \ref{crsec}, we also show the detailed
comparison of the $A'$ differential cross sections calculated at the ETL 
level in this work,  with the one 
calculated  in  Ref.\cite{Liu:2017htz}. The comparison  is made for the case of the aluminium target,
$A'$ masses in the range from 10 keV to 100 MeV and the beam energy $E_0 = 20$ GeV.
In order to better illustrate the comparison between the
two calculations, in  Table \ref{table111} we also show the difference between the values of the total cross sections calculated for
the aluminium target and beam energy 20 GeV using the results of Ref.\cite{Liu:2017htz}.
One can see that for $x$ values in the range $0.5< x <1$ relevant for NA64, the  difference is within $\lesssim 10\%$
for the full mass range. The difference $$\Delta_0=[(d\sigma/dx)_{LM} - (d\sigma/dx)_{GKKK}]/(d\sigma/dx)_{LM}$$ between
the differential cross sections for $x$ values $0.5< x <1$ is varying in the range
$6.5\% \lesssim \Delta_0 \lesssim 9.2\%$ for the
mass range $10$ keV $\lesssim \ma \lesssim 100$ MeV. The observed difference can be interpreted as due to the
accuracy of numerical integration of the $A'$ production
cross section, the figure digitization accuracy \cite{DigitPlot}
and also due to the difference in the nuclei form-factors used in calculations.
It can be conservatively accounted for as additional systematic uncertainty in the evaluation of the $A'$ yield.  

\begin{table}[!htb]
\begin{center}
    \begin{tabular}{|c |c |c |c | }
  \hline
    $m_{A'} $ & $  10$ keV & $ 1 $ MeV & $100$ MeV   \\
    \hline
           \hline
    $\sigma_{tot}^{GKKK}$ [GeV$^{-2}$]  & $2668.78$ & $ 1105.13 $ & $0.212$   \\
    \hline
     $\sigma_{tot}^{LM}$ [GeV$^{-2}$] & $2938.63$ & $1216.97$ & $ 0.230$  \\
    \hline
     $(\sigma_{tot}^{LM} - \sigma_{tot}^{GKKK})/\sigma_{tot}^{LM}$  &
$9.2$ $\%$& $6.5$ $\%$& $7.6$ $\%$ \\
    \hline
    \end{tabular}
 \end{center}
   \caption{   \label{table111} Comparison  of the total cross-sections with those 
 from  Ref.\cite{Liu:2017htz} integrated in the region $0.5<x<1$ for the Al  target, $\epsilon=1$ and $E_0=20$ GeV.}
\end{table}

Independent cross-check has been made by comparing the bremss-spectra obtained with the 
Geant4 package~\cite{Agostinelli:2002hh} and calculated with the integration code. The comparison (see Fig.~\ref{figure7}) shows 
that the relative difference between Geant4 and our calculations does not exceed 2-8 \% for the full bremss. photon energy range.

\subsection{Infrared divergences and applicability of the cross sections  to the mass region $m_{A'} < 1$ MeV.}
As is well known, the infrared divergences arise in some theories (QED, QCD) with massless particles. There are two types 
of infrared singularities - collinear singularities and soft photon(gluon) singularities. In QED with 
nonzero electron mass there are no collinear sinfularities and only soft photon singularities exist. 
To understand the origin of infrared singularities let us consider the expression (6) for the differential 
cross section. For massless electron $m_e = 0$  we have
$ \frac{d\sigma_{{IWW}^{A'}}}{dx}  \sim \frac{1}{(1-x)m^2_{A'}}$ at $1 -x \ll 1$ that corresponds to the collinear singularity at $x =1$.
 Nonzero electron mass 
regulates the collinear infrared singularity. For massless $A'$ boson we have the situation 
with soft dark photon infrared divergences
similar to standard QED, namely 
$\frac{d\sigma_{{IWW}^{A'}}}{dx}   \sim \frac{1}{xm^2_{e}}$ at $x \approx 0 $. So  the  differential cross section is infrared divergent for 
 massless $A'$ in the limit $E_{A~} \rightarrow 0$. Note that according to Bloch-Nordsik theorem in QED with 
finite electron masses it is sufficient to sum over degenerate final states, 
i.e.  soft photons in final states to cancel infrared divergences. For massive $A'$ dark 
photon we have no infrared divergences at all as follows from the expression \eqref{diff-crsec} for the differential 
cross section\footnote{The introduction of nonzero photon mass is often used as infrared regularization in QED.}. Moreover, 
in the search for dark photon $A'$ in NA64 we use rather high cut $E_{A'} >   \frac{E_0}{2} =    50~GeV$, i.e. $x >0.5$\footnote{The use of this 
 cut means that we consider the phase space region only with relativistic $A'$ dark photon}.
\par Let us stress that the  use of this cut means that even   for massless $A'$ boson we
 escape the infrared divergences arising at $x \ll 1$, therefore our calculations for the differential cross sections 
 and the total cross section $\int^{x =1}_{x = 0.5}dx \frac{d\sigma_{{IWW}^{A'}}}{dx}$ are infrared safe and reliable also for the mass
 range $m_{A'} < 1$ MeV. 
Moreover, the standard QED condition $ \frac{\alpha}{\pi} \epsilon^2  \log^2(\frac{t}{m_e^2}) \ll 1 $ for the 
applicability of the perturbation theory works in our case even for $\epsilon = 1$. 

 \begin{figure}[t]
\begin{center}
\includegraphics[width=0.49\textwidth]{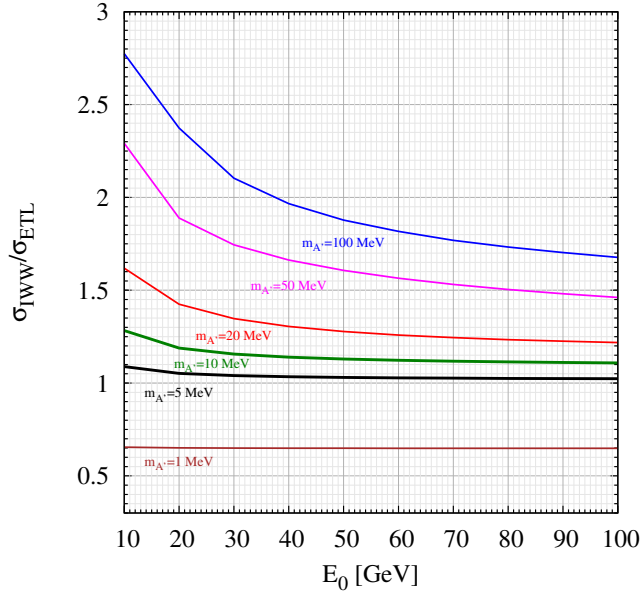}
\caption {The ratio $\sigma_{IWW}/\sigma_{ETL}$ as a function  of 
electron beam energy $E_0$ for various masses $m_{A'}$.
\label{k-factor}}
\end{center}
\end{figure} 

\section{Signal event  simulation}
\subsection{MC simulation of missing energy events.}
First we summarise the  main features of  MC simulation   of the $A'$ 
production  in  NA64. 
The number of $A'$ bosons produced by electrons in the active target is given by 
\begin{equation}
 N_{A'} =\frac{1}{A} \times N_{\text{EOT}} \times \rho \times N_A 
  \times
\sum_{\text{i}} \sigma_{\text{tot}}^{A'} (E_e^i) \times \Delta L_i,  
\label{AprYields}
\end{equation} 
where  $A$ is the atomic weight, $N_A$ is Avogadro's  
number, $N_{\text{EOT}}$ is the number of electrons on target, $\rho$ 
is the target density, $E_e^i$ is the electron energy 
at the $i$th step in the target, $\Delta L_i$ is step length of 
electron path
 and $\sigma_{\text{tot}}^{A'}$ is the 
total cross-section for the $A'$ production  in the  
$e^-Z\rightarrow e^-Z A'$ reaction. 
The summation in \eqref{AprYields} is performed by simulating the electromagnetic shower
development from electrons in 
the active target  with {\tt   GEANT4}.
In the real simulation of signal samples, at each simulation step the following actions are made

\begin{figure}[t]
\begin{center}
\includegraphics[width=0.5\textwidth]{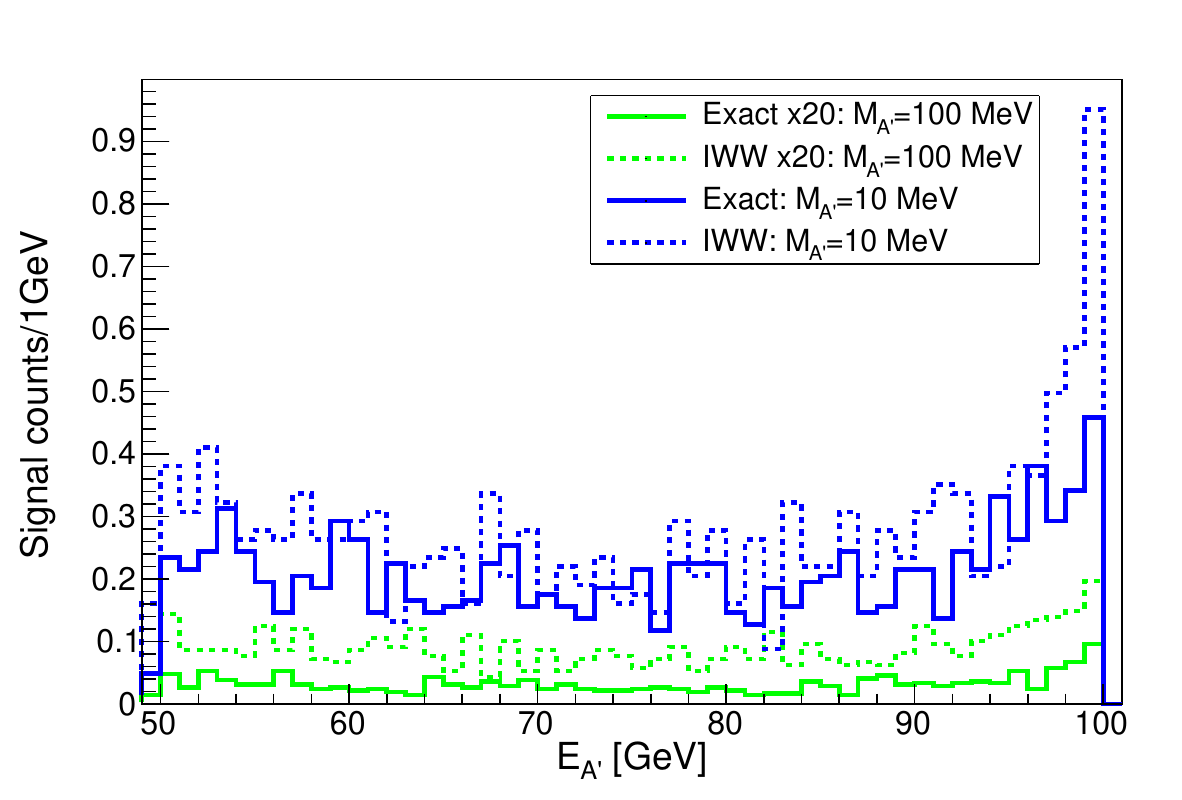}
\caption { Number of signal events at NA64 as a function of 
Dark Photon energy $E_{A'}$ for $m_{A'} = 10$ MeV and  $m_{A'} = 100$ MeV.
\label{FullSimulationSpectraOfDarkPhoton}}
\end{center}
\end{figure}

\begin{itemize}
\item we randomly sample the variable $u_1$ uniformly distributed in the range $[0,1]$,
if $u_1$ is smaller than the $A'$ emission probability 
$$ P_{emission} = \rho N_A \sigma_{\text{tot}}^{A'} \Delta L_i /A,$$
then the emission of $A'$ is accepted,
\item for each emitted $A'$ we randomly generate the values of $x$, 
$\cos \theta$ and the azimuthal angle $\phi$ from the kinematically 
allowed region and then by using the Von Neuman rejection sampling 
we  accept $x$,  $\cos \theta$ and $\phi$ as a signal event values of 
$A'$ emission to calculate 4-momentum of the Dark Photon.
\end{itemize}
Fig.~\ref{FullSimulationSpectraOfDarkPhoton} shows the relevant spectra 
of Dark Photon for Pb target, $\epsilon=10^{-4}$ and $EOT=10^{11}$.
These spectra correspond to missing energy distribution of electron beam  
at NA64. 

\subsection{Simulation with the ETL cross sections}
Now let us describe the  implementation of exact total cross-section 
formula into {\tt {\tt  GEANT4}} numerical simulation of $A'$ production.  
From~\eqref{AprYields} follows that the yield of dark photon is 
proportional to the total cross-section.
Therefore, in order to have the $A'$ yield corresponding to ETL instead of the one calculated with WW approximation in
the {\tt GEANT 4} full simulation
we define the following ratio (K-factor):
\begin{equation}
K(m_{A'},E_0, Z, A) = \sigma^{A'}_{\text{WW}}/\sigma^{A'}_{\text{ETL}},
\label{Rdef1}
\end{equation} 
which represents the overall correction to the 
cross-section (\ref{TotIWWcrSect1}) calculated in the IWW approach.
In Fig.~\ref{k-factor} we show $K$-factors as a 
function of the electron beam energy $E_0$ for various $m_{A'}$.  For heavy nuclei $K$ 
depends rather weakly on $Z$ and $A$, e.g. for the W and Pb nuclei the 
difference is a few $\%$. In our {\tt   GEANT4} MC simulations of the $A'$ 
production we use tabulated K-factors.
From Fig.~\ref{k-factor} and Table 2 one can see that $\sigma_{ETL}$ \
cross-section of $A'$
production is smaller by a factor of $\sim 1.67$ than the corresponding $\sigma_{IWW}$ cross-section obtained 
in IWW approach for $m_{A'}=100$ MeV and $E_0=100$ GeV. On the 
other hand, $\sigma_{ETL}$ exceeds $\sigma_{WW}$ for the Dark Photon 
masses $m_{A'}$ below $5$ MeV. This means that the sensitivity to the mixing strength $\epsilon$~\cite{Gninenko:2016kpg}
for this mass region becomes better.
\begin{table}[!htb]
\begin{center}
    \begin{tabular}{|c |c |c |c |c |c |c|}
  \hline
    $m_{A'} $, MeV & 1 & 5 & $10 $  & $ 20 $ & $ 50 $ & $ 100 $  \\
    \hline  
     $ \sigma_{IWW}/\sigma_{ETL}, \, E_0=100 $ GeV  & 0.648 & 1.023 & 1.10& 1.21& 1.46 &1.67 \\
    \hline
 $\epsilon_{ETL}^2/\epsilon_{IWW}^2 $&0.85&1.14&1.42 &1.78& 1.90 &2.32\\
    \hline     
    \end{tabular}   
 \end{center} 
   \caption{\label{table11}
  The ratio $\sigma_{IWW}/\sigma_{ETL}$ for various masses $m_{A'}$ for the 
electron beam energy $E_0=100$ GeV. The ratio of limits on mixing $\epsilon^2$ 
obtained with a full  MC simulations for the ETL and IWW cross sections, as discussed in Sec. 4, is also shown for different $A'$ masses.  
 }
\end{table}
\begin{figure}[h]
\begin{center}
\includegraphics[width=0.5\textwidth]{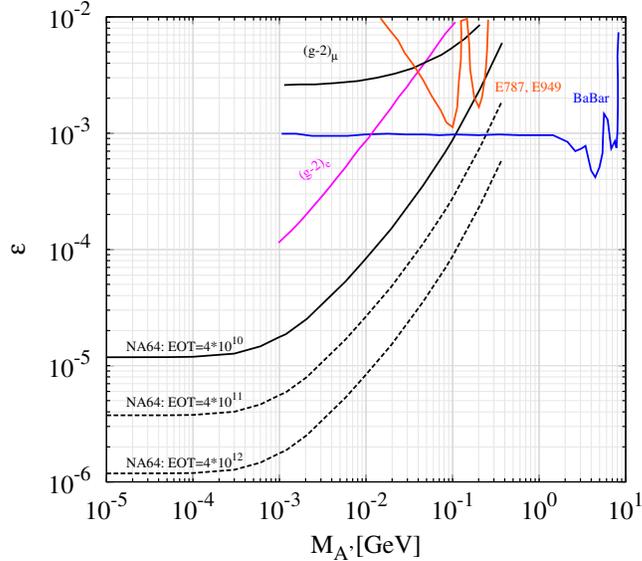}
\caption { Expected upper limits on the mixing $\epsilon$ vs $m_{A'}$  for invisibly decaying $A'$   calculated for the  ECAL target \cite{Andreas:2013lya},
 $E_0 = 100$ GeV ,  and  the missing energy $E_{miss}> 0.5 E_0$ for different numbers of electron on target(EOT).  
Constraints from the    E787 and  E949  \cite{hd,Essig:2013vha} and BaBar \cite{babarg-2} experiments, as well as the muon  $\alpha_\mu$ favored area 
 are also shown. Here, $\alpha_\mu =\frac{g_\mu-2}{2}$,  see  \cite{lightdark1}.
\label{limits}}
\end{center}
\end{figure} 

 \section{ Expected bounds on mixing strength}
In this section we estimate the expected sensitivity of the NA64 experiment to the mixing strength $\epsilon$  
with the {\tt {\tt  GEANT4}} MC simulation.
We calculate the missing energy spectra
of the NA64 and impose the cut $E_{miss}>0.5 E_0$.
By using Eq.(\ref{AprYields}) and Eq.(\ref{Rdef1}) and the relation $n_{A'}^{90\%} > \na$, 
where $n_{A'}^{90\%}=2.3$ is the Poisson 90$\%~CL$ upper limit on the  
number of Dark Photon production events for the background free case and no signal, we obtain 
the expected limits on $m_{A'}$ and $\epsilon$ shown in Fig.~\ref{limits}.
Taking into account discussion in Section 2.3, the bounds are obtained for the mass range $10^{-5}$ - 1 GeV and for 
the total number of 100 GeV  electrons on target
$n_{EOT} = 4\times 10^{10}, 4\times10^{11}$, and $4\times 10^{12}$. It is assumed that the $A'$ decays mainly into invisible final states of
  the Dark sector, $A'\rightarrow \chi \chi$.
The obtained results show that the expected sensitivity of the NA64 experiment potentially allows to test the still unexplored area of the mixing
strength  $10^{-6}\lesssim \epsilon \lesssim 10^{-2}$ and masses from $m_{A'} \lesssim 1$ GeV up to a very small values $\ll 1$~MeV.

\section{Summary}
Here we briefly summarise  the main improvements achieved in this work with respect to our previous paper \cite{Gninenko:2016kpg}, 
as well as the recent work on cross section calculations carried out in Ref.\cite{Liu:2017htz}. 
We have studied  the missing energy signature of the dark photon production in the  reaction \eqref{reaction} 
   in the experiment NA64 at the CERN SPS.  For this study we performed the exact tree-level calculations of the  $A'$ production cross sections and implemented 
   them  into the {\tt GEANT4} based simulation program  of the NA64 detector. 
   We cross-checked our results with the results of Ref.\cite{Liu:2017htz} and found that they are in a quite good agreement for the
   wide mass range.  We believe that the error of the estimate of the  total cross sections in the sub-MeV - sub-GeV mass range   
obtained in these two works hardly exceeds a few \%. This small difference can be used as a systematic uncertainty of the $A'$ yield. 
For a realistic study of the expected  sensitivity of the NA64 experiment we made the following improvements:
we implemented the ETL $A'$ cross sections in the {\tt  GEANT4} simulation package and 
performed the full simulation of the detector response to the  $\ainv$ signal events in a wide mass region.
We have evaluated the expected sensitivity of the NA64 experiment and have shown  that it potentially allows to probe the still
unexplored area on the $ m_{A'}, \epsilon$ plane for $10^{-6}\lesssim \epsilon \lesssim 10^{-2}$ and a wide $A'$ mass range
including the sub-MeV region.
 \vskip0.5cm
{\large \bf Acknowledgments}

We would like to thank the members of the NA64 Collaboration for numerous
discussions. D.K. would like to thank I. Karpikov and K. Astapov for fruitful discussion.
The work of D.K. on simulations of signal events has been supported by the RSF grant 14-22-00161.

\end{document}